# Impact of Mechanical Stress on IGZO TFTs: Enhancing PBTI Degradation

K. Vishwakarma[1], K. Lee[1*], A. Kruv[1], A. Chasin[1], M. J. van Setten[1], C. Pashartis[1], O. O. Okudur[1], M. Gonzalez[1], N. Rassoul[1], A. Belmonte[1], B. Kaczer[1]

[1] imec, Kapeldreef 75, 3001, Leuven, Belgium
kavita.vishwakarma@imec.be

*now with Intel Hillsboro, US

*Abstract*—This study investigates the impact of out-of-plane compressive mechanical stress (MS) on the performance and reliability of n-channel IGZO thin film transistors (TFTs). It is demonstrated that MS induces a positive $V_{th}$ shift in the device transfer characteristics and enhances electron trapping during Positive Bias Temperature Instability (PBTI) tests. These effects are attributed to the widening of the IGZO bandgap ($E_G$) and increased accessibility of carriers to $AlO_X$ gate oxide trap levels. As substantial residual MS is generated in 3D device processing, understanding its impact on IGZO TFTs is crucial for enabling future 3D DRAM technology.

## I. INTRODUCTION

The key advantages of IGZO-TFT such as low-temperature processing, BEOL compatibility, reasonable mobility, and high transparency have led to them being employed in drivers of displays [1], flexible electronics [2], and in RF applications [3]. Moreover, the ultra-low leakage current of IGZO TFT (down to $\sim 10^{-22}$ A/µm) [4], which results in its long retention, has positioned IGZO as a strong candidate for DRAM applications, enabling a capacitorless (2T0C) configuration [5]. It is expected that 3D integration will be adopted in the future to enable further DRAM scaling, similar to what has been done in 3D NAND [6].

3D integration relies on processing high aspect ratio multi-material stacks, thermal treatments, and packaging [7], [8]. During these processes, the mismatch in thermal expansions coefficients and lattice constants of the stacked materials leads to build-up of residual Mechanical Stress (MS), ranging from hundreds of MPa to a few GPa [9], [10], which affects electrical performance and reliability. While it is generally known that MS influences electrical characteristics of semiconductor devices, such as Field Effect Transistors (FETs), FinFETs, Ferroelectric RAM (FeRAM), and 3D NAND [11] – [14], there have been a very few reports of the MS impact on IGZO, limited to theoretical calculations [15] and tests involving IGZO TFTs on flexible substrates [16], [17]. However, such methods induce stress globally across the substrate, and the MS levels generated through these bending techniques are comparatively low [18].

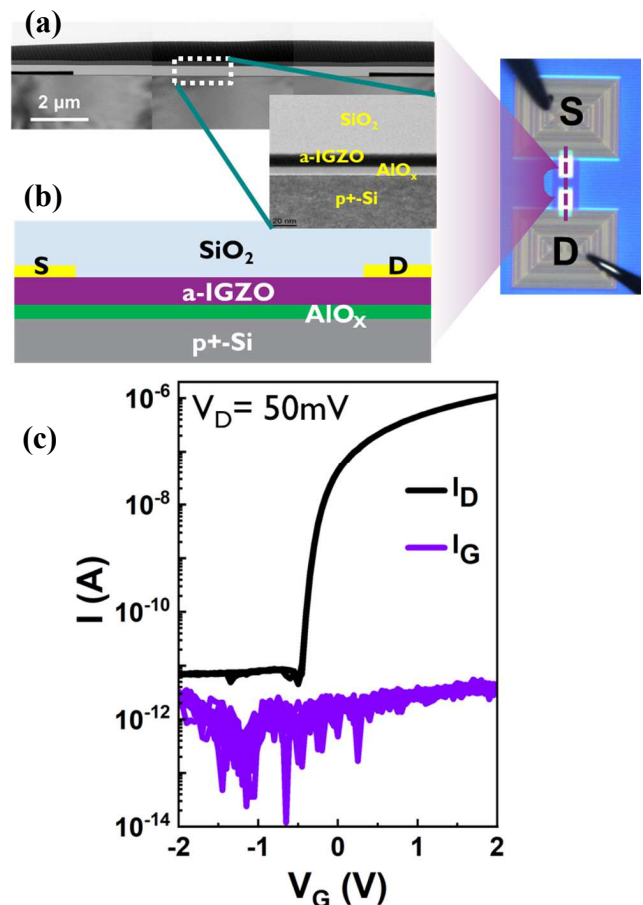

**Fig. 1.** (a) Cross-sectional TEM image through a 10 × 10 µm² IGZO device showing its distinctive layers. (b) Device schematic with IGZO (12 nm) and $AlO_X$ (10 nm). (c) Pre-stress (time - zero) $I_D$-$V_G$ and $I_G$-$V_G$ curves at $V_D$= 50 mV, demonstrating good device behavior.

This study provides insights into the impact of *locally* applied ~ GPa level MS on IGZO TFTs fabricated in a 300 mm

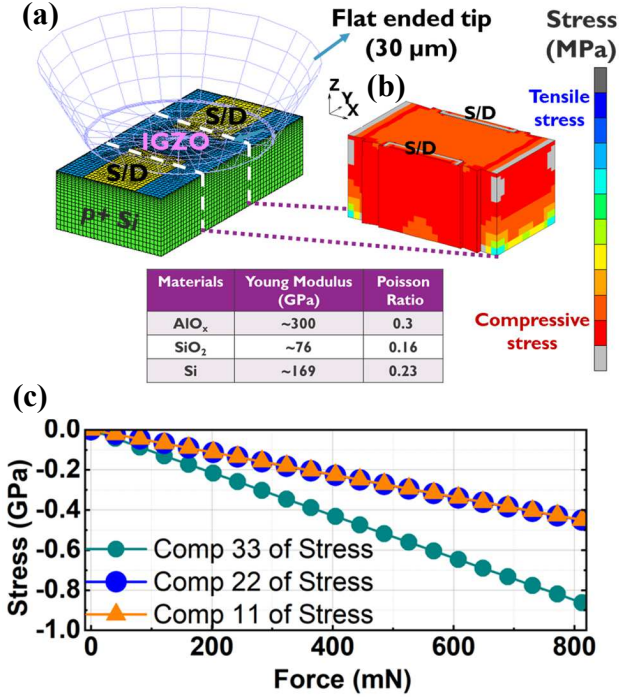

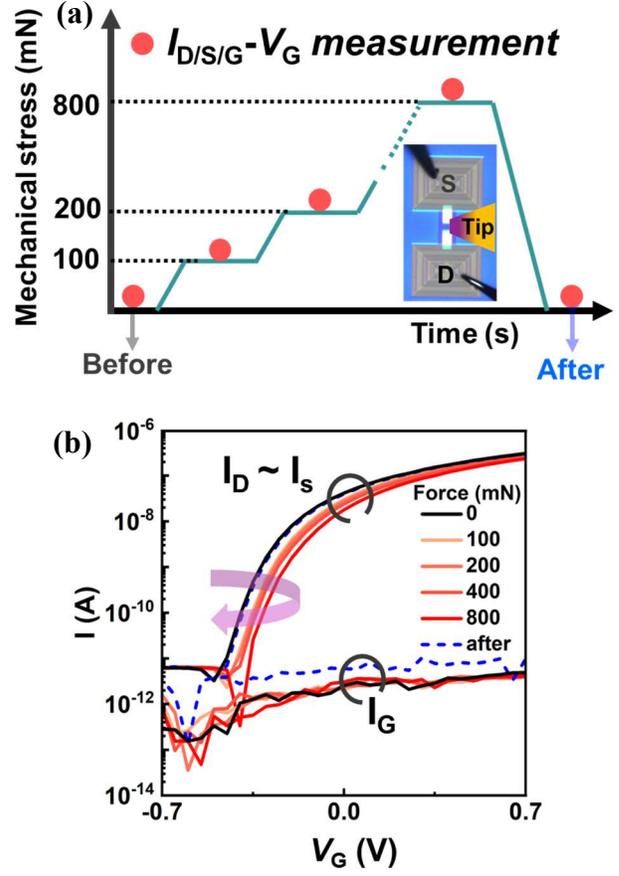

**Fig 2**. (a) 3D FEM model with 30 μm flat-ended tip with model parameters. The IGZO parameter (elastic tensor components) was extracted from DFT computations. (b) Stress distribution under 800 mN tip load shows uniform compressive stress in the channel region, and (c) converted reaction force into stress values (x, y, z directions) at channel centre, highlighting higher -z direction (comp 33) stress of ~0.9 GPa at 800 mN.

**Fig. 3.** (a) Schematic of the applied mechanical force versus time on the device (inset), with I-V measurements taken at each force levels. (b) I-V curves before (0 mN), during (100-800 mN), and after the application of force. The after force I-V curve overlaps with 0 mN indicating complete recovery. A positive shift was observed with increasing force, while gate leakage remained stable up to 800 mN.

FAB. The *locally* applied external force at the ~GPa level of MS was introduced using a nanoindenter (NI), which enables precise control over both the amplitude of the force and the exact location of its application. We first investigate the effects of MS on the "time-zero" performance of IGZO-TFTs, observing full recovery in its characteristics after stress. Additionally, since IGZO is inherently an n-channel material, PBTI reliability is identified as the primary reliability concern to be studied [19]. Consequently, the combined impact of MS and electrical stress (ES) as PBTI, resulting in enhanced degradation of IGZO-TFTs, is also investigated. Our findings support the understanding that PBTI degradation at room temperature is primarily caused by electron trapping in the gate oxide [20], which is further intensified by MS, a phenomenon not previously studied in IGZO under MS.

## II. EXPERIMENTAL

The tests were performed on back-gated IGZO-TFTs with dimensions of W × L = 10 × 10 μm², as shown in Figs. 1 (a) & (b). The devices were processed on a 300 mm p+-Si substrate, featuring a 10 nm $AlO_X$ gate dielectric, a 12 nm amorphous PVD $In_{38}Ga_{38}Zn_{24}O_X$ channel, and TiN/W source/drain (S/D) contacts. Before applying MS, electrical tests (at time zero) were performed [20], with the substrate serving as the back gate. Figure 1(c) shows the measured pre-stress (time-zero) transfer curves over multiple times, demonstrating stability with no change in characteristics and low gate leakage at a drain bias of 50 mV.

For performing MS test only, an external force (100-800 mN) was applied in steps on top of device channel or S/D contacts using TI Hysitron 950 NI equipped with a flat non-conducting tip of 30 μm diameter. The externally applied force was converted to MS value using finite-element modelling (FEM) in Hexagon Marc software [21]. The model and its parameters are shown in Fig. 2(a). The applied force induced triaxial compressive stress in the channel region (Fig. 2 (b)), with the vertical stress (comp 33) being the dominant component reaching up to ~0.9 GPa at 800 mN, as can be observed from Fig. 2(c). $I_{D/S/G}$-$V_G$ curves were measured at each force level, including both before- and after-force applications, as shown in Fig. 3(a). The impact of MS on the device performance was evaluated by assessing the threshold voltage ($V_{th}$), on-current ($I_{ON}$), and the subthreshold swing (SS). The shift in threshold voltage ($\Delta V_{th}$) relative to $V_{th0}$ was extracted from the $I_D$-$V_G$ curves at a fixed current of ~5 nA.

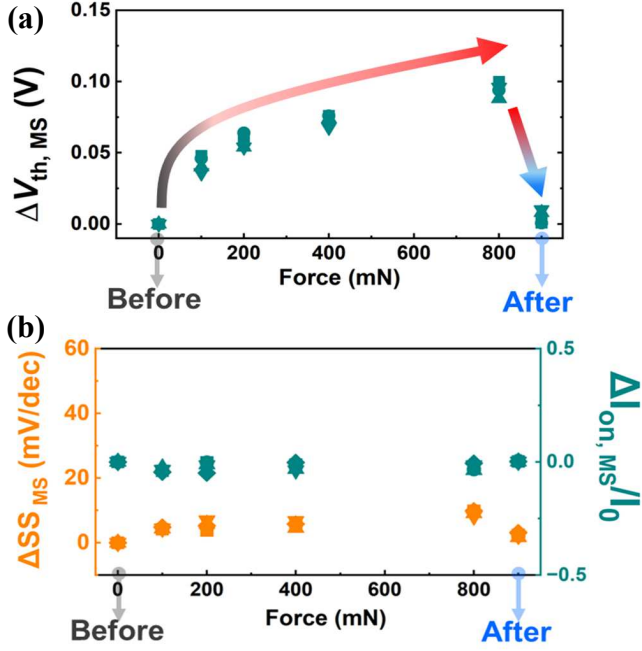

**Fig. 4.** (a) Shift in $\Delta V_{th, MS}$ versus force, indicating that the +ve shift increases with applied force (b) demonstrating minimal diminishment in SS and $I_{ON}$ under the influence of force.

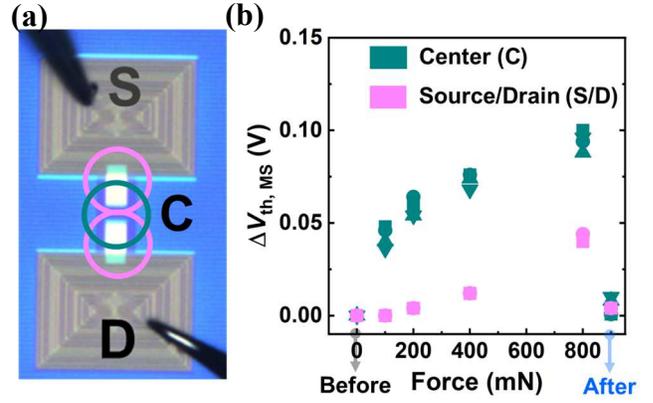

**Fig. 5.** (a) Schematic of tip position variation over Source (S), Center (C), and Drain (D). (b) $\Delta V_{th, MS}$ shift vs force, showing a greater impact at C compared to S/D, as MS over channel reduced as indented far from the device centre.

The subthreshold Swing (SS) was extracted between the points corresponding to 0.3 and 0.03 nA, while the on-current ($I_{ON}$) was extracted at $V_{th}$+1V.

The Electrical Stress (ES) test, as PBTI, was performed by applying a constant positive gate bias from 2 to 5 V. Simultaneously, an external force of 800 mN was applied over the channel region as the MS+ES test, and $I_{D/S/G}$-V curves were recorded periodically for a total stress time ($t_{stress}$) of 500 seconds. The impact of MS on the TFTs and their PBTI has been studied only at room temperature, as the NI system does not currently allow changing the temperature of the measured device.

### III. RESULTS AND DISUSSIONS

Following the stable behavior of TFTs at time zero, the response of the application of only MS, as well as the combination of both MS and ES on the TFTs, will be discussed further in the subsections below.

#### A. Application of MS only

Application of MS above the device channel region resulted in a positive $V_{th}$ shift ($\Delta V_{th, MS}$ up to ~100 mV at 800 mN), as shown in Fig. 3 (b). Whereas the SS, $I_{ON}$, and $I_G$ were almost unaffected, as can be seen from the $I_D$-$V_G$ curves (Fig. 3 (b)) and the extracted values in Figs. 4 (a) – (b). The device performance fully recovered upon the force removal, indicating no permanent (plastic) deformation. The observed $V_{th}$ shift is caused by the force application and not by the device instability, as the device performance stayed stable after 10 repeated measurements in the absence of applied force, previously shown in Fig. 1(c). When the tip was moved from the centre to the either S or D regions, leading to lower MS in the channel and higher MS at the contacts (Figs. 5 (a) , (b)), $\Delta V_{th, MS}$ exhibited minimal change. This indicates that the observed effect is channel- rather than contacts-limited.

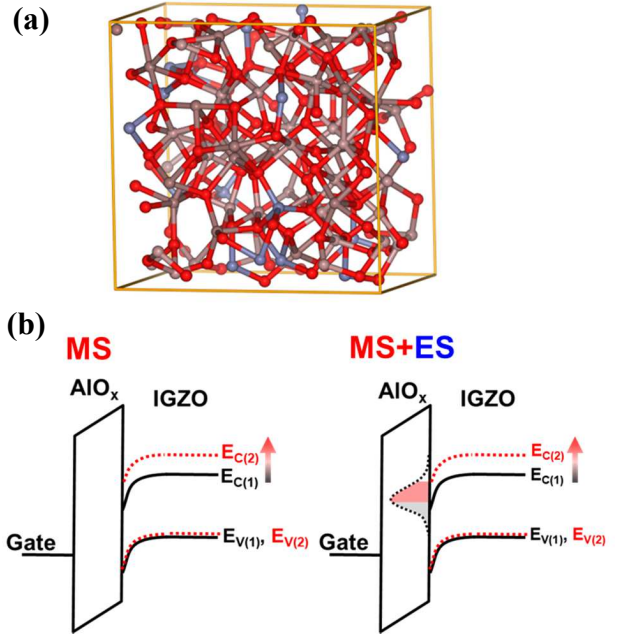

**Fig. 6.** (a) The 200 atom unit cell of amorphous IGZO used for the computation of the elastic properties and band energy dependence on strain. Computations performed using the PBEsol exchange correlation functional. (b) An energy band diagram shows the IGZO conduction ($E_C$) and valence ($E_V$) bands, labelled $E_{C(1)}$ and $E_{V(1)}$ without MS, and $E_{C(2)}$ and $E_{V(2)}$ with MS. Applying MS causes a +ve shift in $\Delta V_{th}$, likely due to a ~50 meV shift from $E_{C1}$ to $E_{C2}$. With simultaneous MS+ES, $E_C$ shifts align more with the $AlO_X$ trap band, increasing trapping probability.

From density functional theory (DFT) computations following the methodology described in [22], on structural models (Fig. 6 (a)) of amorphous stoichiometric IGZO ($InGaZnO_4$) as developed in [23]. we expect that the applied force (800 mN) induces a strain of -0.004 (mm/mm), leading to an increase in the conduction band ($E_C$) level by approximately 50 meV. The shift in the $E_C$ level with MS is visualized in the energy band-diagram sketch (Fig. 6 (b)). Therefore, the observed positive $\Delta V_{th, MS}$ can be attributed to the widened $E_G$ (reduced electron affinity) and to MS-enhanced electron trapping in the gate dielectric. This will further be discussed with respect to PBTI mechanism in the next section.

## B. Application of both MS and ES

Fig. 7 illustrates the experimental method used to evaluate the impact of MS on device PBTI, where an 800 mN force and ES were applied simultaneously during the stress time ($t_{stress}$), with I-V curves collected periodically. Application of MS combined with ES resulted in the enhancement of electron trapping in the gate oxide. In the absence of MS (ES only), $I_D$-$V_G$ curves show a parallel positive $V_{TH}$ shift as the result of ES ($V_{G,str}$=3V and 5V), indicating the primary mechanism being electron trapping in $AlO_X$, Figs. 8 (a, c). The combination of MS+ES intensified PBTI degradation, see Figs. 8 (b, d), leading to stronger positive $V_{th}$ shift. Fig. 9 (a) exhibits $\Delta V_{th}$ shift vs $t_{stress}$ plot with enhanced degradation on MS+ES with the method of application schematically shown below in Fig. 9 (b). The clear enhancement in degradation can be observed in Fig. 10 (a), where $\Delta V_{th}$ shift @ $t_{stress}$=1 s for 800 mN was subtracted for the remaining $t_{stress}$ values. Both ES and MS+ES approximately follow a power-law kinetic behaviour ($\Delta V_{th} \sim t^n$), as shown in Figs. 11(a) and (b). A smaller exponent 'n' for MS+ES suggests an enhanced trapping process [24]. Additionally, the voltage acceleration factor ($\gamma$) was determined from the power-law dependence on overdrive voltage ($V_{OV}$) at $t_{stress}$=500 s (Fig. 12). A higher '$\gamma$' during ES only, indicates a favourable energy misalignment between the AlO$_X$ trap band and IGZO energy conduction band ($E_C$) [24]. Under MS+ES, the misalignment decreases, as illustrated in the band diagram (Fig. 6 (b)), leading to a +$\Delta V_{th, MS+ES}$ shift and reduced 'γ' (Fig. 12) due to the greater accessibility of trap bands.

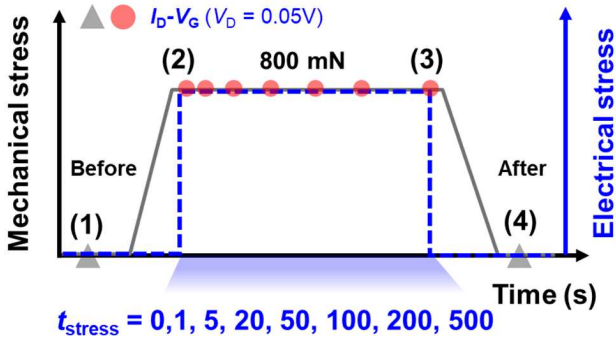

**Fig. 7.** Schematic showing MS of 800 mN (black) and simultaneous ES (blue) methodology versus time, with I-V measurements at stress times ($t_{stress}$) from 1 to 500 s.

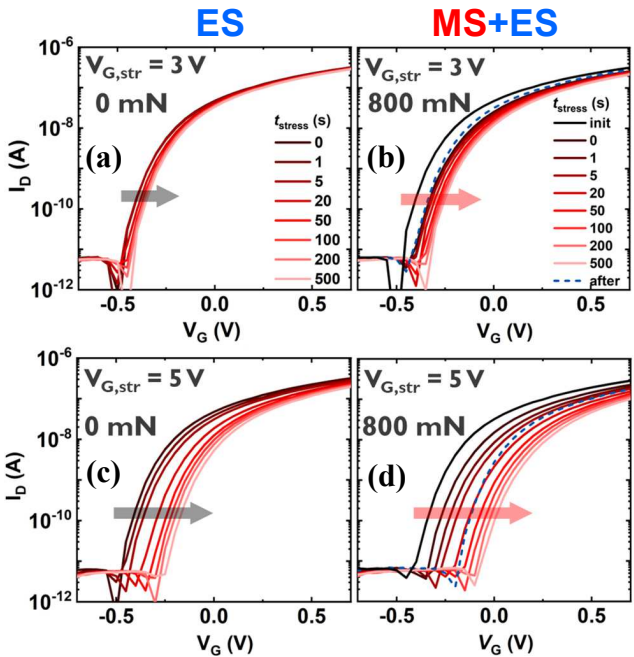

**Fig. 8.** (a) and (c) An ES measurement with gate terminal stress ($V_{G,str}$) of 3V and 5V, shows positive parallel I-V shifts at $V_D$=50 mV. (b) and (d) MS with simultaneous ES enhances the positive shift of I-V without degrading $I_{ON}$ and SS.

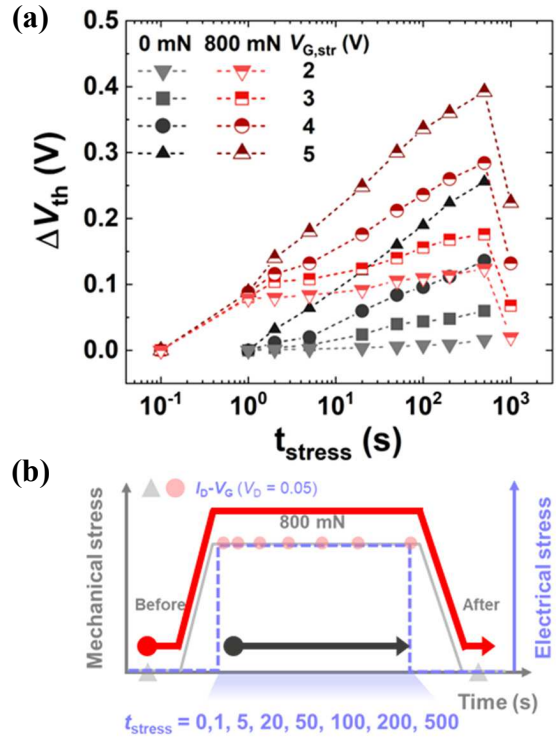

**Fig. 9.** (a) $\Delta V_{th}$ is extracted from the full I-V curves at fixed current level of ~ 5 nA for $V_{G,str}$ from 2 to 5V. Extracted $\Delta V_{th}$ shift vs $t_{stress}$ plot for both ES (black) and ES+MS (red) over the whole range from before to after force as method schematically shown below in (b).

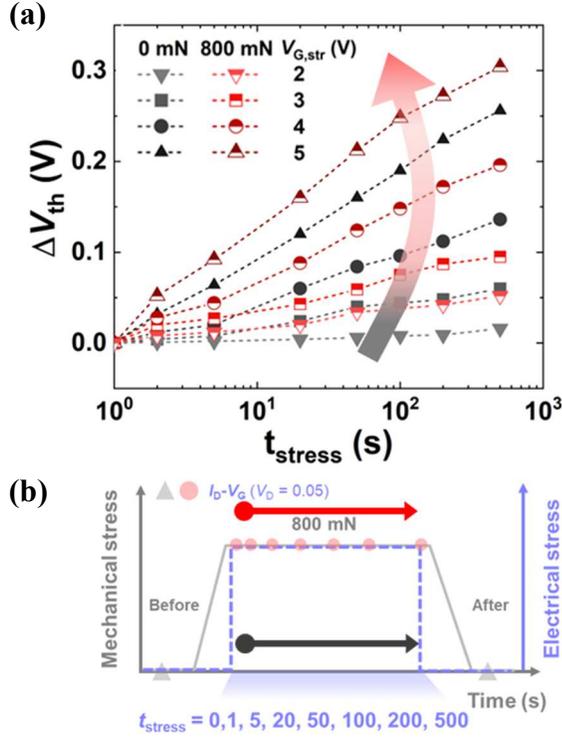

**Fig 10.** (a) For proper comparison, the $\Delta V_{th}$ shift @ t=1 s for 800 mN was subtracted for the remaining $t_{stress}$, signifies clear enhancement in degradation from only ES with its method shown in (b).

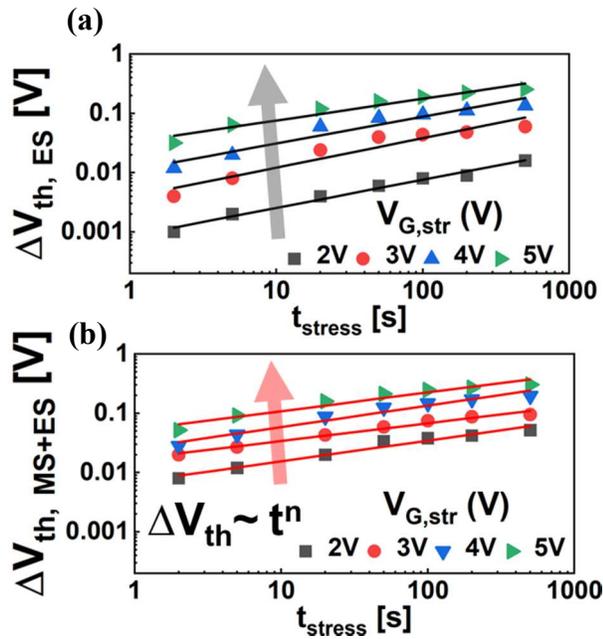

**Fig. 11.** Degradation kinetics (time exponent ('n')) changes as a function of applied $V_{G,str}$ : time exponent decreases as a function of $V_{G,str}$. (a) for ES based measurement. (b) for MS+ES based measurement The 'n' is smaller for MS+ES, signifying the AlOx trap band being more accessible.

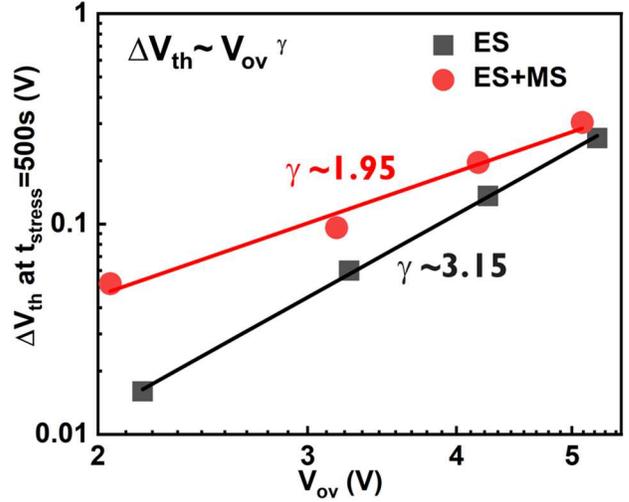

**Fig. 12.** The power law dependence over overdrive voltage ($V_{OV}$), with exponent 'γ' as a voltage acceleration factor. Decrement of 'γ' exhibits the strong effect of MS+ES in comparison to ES only.

## IV. CONCLUSIONS

We demonstrated that IGZO-TFTs subjected to MS exhibit a positive shift in transfer characteristics, which fully recovers after MS is removed, with minimal influence on SS and $I_{ON}$. This behavior is likely associated with widening of IGZO $E_G$, enhanced electron trapping in AlO$_X$ or IGZO, and change in AlO$_X$ trap energy levels under ~ GPa stress. These changes enhance trapping probability by reducing the misalignment between the trap band and the IGZO conduction band, as sufficed by the decrease in 'γ', thereby accelerating degradation during PBTI. This study provides valuable insights for optimizing the integration process of IGZO-TFTs in 3D DRAM configurations.

## ACKNOWLEDGMENT

*K. Vishwakarma acknowledges the funding from the European Union (EU)'s Horizon Europe Framework Programme under the Marie Skłowdowska-Curie (MSCA) Postdoctoral Fellowship Action 2023 – MINDSET with grant agreement No. 101154357 from the European Commission. This work was supported by the industry affiliated Active Memory Program and This work has been enabled in part by the NanoIC pilot line. The acquisition and operation are jointly funded by the Chips Joint Undertaking, through the European Union's Digital Europe (101183266) and Horizon Europe programs (101183277), as well as by the participating states Belgium (Flanders), France, Germany, Finland, Ireland and Romania*

REFERENCES

[1] H.-J. Shin, S. Takasugi, W.-S. Choi, M.-K. Chang, J.-Y. Choi, S.-K. Jeon, S.-H. Yun, H.-W. Park, J.-M. Kim, H.-S. Kim, C.-H. Oh, "A Novel OLED Display Panel with High-Reliability Integrated Gate Driver Circuit using IGZO TFTs for Large-Sized UHD TVs," *SID Symp. Dig. Tech. Papers*, vol. 49, no. 1, pp. 358–361, 2018.

[2] K. Myny, "The development of flexible integrated circuits based on thin-film transistors", *Nature Electronics*, vol.1, p.30-39, 2018.

[3] A. Chasin, V. Volskiy, M. Libois, K. Myny, M. Nag, M. Rockelé, G. A. E. Vandenbosch, J. Genoe, G. Gielen, and P. Heremans "An Integrated a-IGZO UHF Energy Harvester for Passive RFID Tags," *IEEE Transactions on Electron Devices*, vol. 61, no. 9, pp. 3289-3295, Sept. 2014.

[4] H. Kunitake *et al.*, "A c-Axis-Aligned Crystalline In-Ga-Zn Oxide FET With a Gate Length of 21 nm Suitable for Memory Applications," *IEEE Journal of the Electron Devices Society*, vol. 7, pp. 495-502, 2019.

[5] A. Belmonte, H. Oh, N. Rassoul, G.L. Donadio, J. Mitard, H. Dekkers, R. Delhougne, S. Subhechha, A. Chasin, M. J. van Setten, L. Kljucar, M. Mao, H. Puliyalil, M. Pak, L. Teugels, D. Tsvetanova, K. Banerjee, L. Souriau, Z. Tokei, L. Goux, G. S. Kar, "Capacitor-less, Long-Retention (>400s) DRAM Cell Paving the Way towards Low-Power and High-Density Monolithic 3D DRAM," *in IEEE International Electron Devices Meeting (IEDM)*, San Francisco, CA, USA, pp. 28.2.1-28.2.4, 2020.

[6] H. Tanaka, M.Kido, K.Yahashi, M.Oomura, R.Katsumata, M.Kito, Y.Fukuzumi, M.Sato, Y.Nagata, Y.Matsuoka, Y.Iwata, H.Aochi and A.Nitayama, "Bit Cost Scalable Technology with Punch and Plug Process for Ultra High Density Flash Memory," *2007 IEEE Symposium on VLSI Technology*, Kyoto, Japan, pp. 14-15, 2007.

[7] S.-H. Lee, "Technology scaling challenges and opportunities of memory devices," *IEEE International Electron Devices Meeting (IEDM)*, San Francisco, CA, USA, pp. 1.1.1-1.1.8, 2016.

[8] K. B. Kim, Y. T. Oh, and Y. H. Song, "Simulation of residual stress and its impact on a poly-silicon channel for three-dimensional, stacked, vertical-NAND flash memories," *Journal of the Korean Physical Society*, vol. 70, no. 12, pp. 1041–1048, 2017.

[9] D. Yoon, J. Sim, and Y. Song, "Mechanical stress in a tapered channel hole of 3D NAND flash memory," *Microelectronics Reliability*, vol. 143, p. 114941, 2023.

[10] J. Lee, D. G. Yoon, J. M. Sim, and Y. H. Song, "Impact of residual stress on a polysilicon channel in scaled 3D NAND flash memory," *Electronics*, vol. 10, no. 21, p. 2632, 2021.

[11] K. Lee, B. Kaczer, A. Kruv, M. Gonzalez, R. Degraeve, S. Tyaginov, and I. De Wolf, "Hot-electron-induced punch-through (HEIP) effect in p-MOSFET enhanced by mechanical stress," *IEEE Electron Device Letters*, vol. 42, no. 10, pp. 1424–1427, 2021.

[12] T. Furuhashi, M. Haneda, T. Sasaki, Y. Kagawa, Y. Ooka, T. Hirano, et al., "Characterization of impact of vertical stress on FinFETs," in *2019 22nd European Microelectronics and Packaging Conference & Exhibition (EMPC)*, pp. 1–4, Sep. 2019.

[13] Y. Liu, S. Clima, G. Hiblot, P. Matagne, M. L. Popovici, B. Kaczer, and I. De Wolf, "Investigation of the impact of externally applied out-of-plane stress on ferroelectric FET," *IEEE Electron Device Letters*, vol. 42, no. 2, pp. 264–267, 2021.

[14] A. Kruv, A. Arreghini, D. Verreck, M. Gonzalez, I. De Wolf, and M. Rosmeulen, "Impact of mechanical stress on 3-D NAND flash current conduction," *IEEE Transactions on Electron Devices*, vol. 67, no. 11, pp. 4891–4896, Nov. 2020.

[15] A. D. J. De Meux, G. Pourtois, J. Genoe, and P. Heremans, "Comparison of the electronic structure of amorphous versus crystalline indium gallium zinc oxide semiconductor: structure, tail states and strain effects," *Journal of Physics D: Applied Physics*, vol. 48, no. 43, p. 435104, 2015.

[16] M. M. Billah, J. U. Han, M. M. Hasan, and J. Jang, "Reduced mechanical strain in bendable a-IGZO TFTs under dual-gate driving," *IEEE Electron Device Letters*, vol. 39, no. 6, pp. 835–838, 2018.

[17] M. M. Billah, M. M. Hasan, and J. Jang, "Effect of tensile and compressive bending stress on electrical performance of flexible a-IGZO TFTs," *IEEE Electron Device Letters*, vol. 38, no. 7, pp. 890–893, 2017.

[18] A. Kruv, B. Kaczer, A. Grill, M. Gonzalez, J. Franco, D. Linten, and I. De Wolf, "On the impact of mechanical stress on gate oxide trapping," *in IEEE International Reliability Physics Symposium (IRPS)*, pp. 1-5, 2020.

[19] P. Rinaudo, A. Chasin, J. Franco, Z. Wu, S. Subhechha, G. Arutchelvan, G. Eneman, B. Y. V. Ramana, N. Rassoul, R. Delhougne, B. Kaczer, I. D. Wolf, and G. S. Kar, "Degradation Mapping and Impact of Device Dimension on IGZO TFTs BTI," *IEEE Transactions on Device and Materials Reliability*, vol. 23, no. 3, pp. 337-345, Sept. 2023.

[20] A. Chasin, J. Franco, K. Triantopoulos, H. Dekkers, N. Rassoul, A. Belmonte, and G. S. Kar, "Understanding and modelling the PBTI reliability of thin-film IGZO transistors," in *IEEE International Electron Devices Meeting (IEDM)*, pp. 31-1, Dec. 2021.

[21] https://hexagon.com/products/marc.

[22] C. Pashartis, M.J. van Setten, M. Houssa, and G. Pourtois, "Computing elastic tensors of amorphous materials from first-principles," *Computational Materials Science*, vol. 242, no. 113042, Jun. 2024.

[23] A. Kruv, M.J. van Setten, A. Chasin, D. Matsubayashi, H.F.W. Dekkers, A. Pavel, Y. Wan, K. Trivedi, N. Rassoul, J. Li, Y. Jiang, S. Subhechha, G. Pourtois, A. Belmonte, and G. Sankar Kar, "In-poor IGZO: superior resilience to hydrogen in forming gas anneal and PBTI", arXiv preprint arXiv:2412.07362.

[24] J. Franco, V. Putcha, A. Vais, S. Sioncke, N. Waldron, D. Zhou, and B. Kaczer, "Characterization of oxide defects in InGaAs MOS gate stacks for high-mobility n-channel MOSFETs," in *IEEE International Electron Devices Meeting (IEDM)*, pp. 7-5, 2017.